\documentclass[12pt]{iopart}
\usepackage{dcolumn}
\usepackage{bm}
\expandafter\let\csname equation*\endcsname\relax
\expandafter\let\csname endequation*\endcsname\relax
\usepackage{amsmath}
\usepackage{amsfonts}
\usepackage{amssymb}
\usepackage{array}
\usepackage{graphicx}
\usepackage{textcomp}

\begin{document}
\title{Quantum teleportation of spin coherent states: beyond continuous variables teleportation}
\author{Alexey N. Pyrkov$^{1,2}$ and Tim Byrnes$^{1}$}
\address{$^1$ National Institute of Informatics, 2-1-2
Hitotsubashi, Chiyoda-ku, Tokyo 101-8430, Japan}
\address{$^2$ Institute of Problems of Chemical Physics RAS, Acad. Semenov av., 1, Chernogolovka, 142432, Russia}
\begin{abstract}
We introduce a quantum teleportation scheme that transfers a spin coherent state between two locations using entanglement. In the scheme an unknown spin coherent state lying on the equator of the Bloch sphere, such as realized in a coherent two-component Bose-Einstein condensate, is teleported onto a distant spin coherent state using only elementary operations and measurements. The scheme works in the regime beyond the standard continuous variables approximation based on the Holstein-Primakoff transformation. We analyze the error of the protocol with the number of particles $ N $ in the spin coherent state under decoherence and find that it scales favorably with $ N $. The simplicity of the operations involved and the robustness under decoherence should make the protocol suitable under realistic experimental conditions.
\end{abstract}
\maketitle
\section{Introduction}
Quantum mechanics is typically associated with the microscopic world, where phenomena such as superposition
and entanglement occur for few-particle systems, but become difficult to observe when scaled up to 
macroscopic objects. It has however become clear in recent years that even macroscopic objects 
can behave quantum mechanically, and possess significant amounts of entanglement \cite{vedral08,bravyi12}. 
The quantum-to-classical transition occurs not inherently 
because of the large number of particles, but due to fast decoherence of such macroscopic objects \cite{schlosshauer07,zurek03}. 
Nevertheless, observing and manipulating quantum mechanics beyond the few-qubit level remains a serious challenge in the field of quantum technologies. 

Recently there have been several remarkable experimental demonstrations of entanglement generation and teleportation 
between macroscopic objects \cite{fickler12,sherson06,julsgaard01,chou05,krauter12,bao12}. 
These follow from the many successful realizations of teleportation using a variety of systems ranging from those using photons \cite{bouwmeester97,boschi98,furusawa98}, atoms \cite{riebe04,barrett04,olmschenk09}, and hybrid systems \cite{sherson06,chen08}. 
In Ref. \cite{julsgaard01} spin polarized atomic ensembles were used to form continuous variables (CV) \cite{braunstein05} and entanglement was generated between two ensembles. In Ref. \cite{krauter12} such CV variables were teleported between two atomic ensembles, while
in Ref. \cite{bao12} a spin wave qubit state was teleported. Although these demonstrations use macroscopic systems, in terms of the teleportation protocol they fall into one of two categories - qubit \cite{bennett93} or CV teleportation~\cite{vaidman94}. In particular for CV teleportation using ensembles, typically the spins are polarized in the $ S^x $-direction, and the remaining spin directions $ S^y, S^z $ (suitably normalized) form canonical position and momentum quadratures \cite{braunstein05}.  The teleported states are then small displacements from the $ S^x $-polarized state on the Bloch sphere \cite{krauter12}. For spin coherent states that can deviate significantly from a given polarized state, the CV teleportation schemes breaks down, and there is no simple scheme to teleport an arbitrary spin coherent state on the Bloch sphere. Here we introduce a protocol for teleporting a spin coherent state that has an arbitrary position on the equator of the Bloch sphere.

Previously one of the authors proposed an alternative way of performing quantum computation, involving spin coherent states \cite{byrnes12}. The method encodes qubit information on spin coherent states, which can involve 
either spinor Bose-Einstein condensates (BECs) or atomic ensembles. Taking the example of the two-component BEC,
a single qubit is encoded as
\begin{equation}
\label{becqubit}
|\alpha,\beta\rangle\rangle\equiv\frac{1}{\sqrt{N!}}(\alpha a^\dagger+\beta b^\dagger)^{N}|0\rangle,
\end{equation}
where creation operators for the two hyperfine states $ a^\dagger, b^\dagger $ obey bosonic commutation relations, 
and $ N $ is the number of bosons in the BEC. The concept of the work of Ref. \cite{byrnes12} is that 
such ``BEC qubits'' can be used with analogous properties to standard qubits, such that 
various quantum algorithms and quantum communication \cite{pyrkov12} can be performed using these states. Although such spin coherent states are
generally thought to be ``classical'' in the sense that the fluctuations of normalized spin variables
$ S^j/N $ decay as $ \propto \frac{1}{\sqrt{N}} $ ($j=x,y,z$) \cite{gross12}, it was shown that quantum effects 
such as entanglement can be generated between several such BEC qubits \cite{byrnes13}, and quantum 
algorithms could be performed provided an equivalent algorithm could be found.

It is clear from the structure of (\ref{becqubit}) that the quantum information is duplicated $ N $ times
when encoded in this way. Such a structure is attractive from a quantum technology standpoint since it 
adds a robustness via duplication of the information. Thus unlike single qubit systems (not 
involving quantum error-correction) where a single quanta of external noise can destroy the quantum 
information, even a loss of some fraction of the number of particles in (\ref{becqubit}) does not
result in complete loss of the quantum information, it merely contributes to a diminished signal amplitude.
The question is then whether such states (\ref{becqubit}) can indeed 
be used to perform quantum information processing tasks, as the Hilbert space is complicated by the 
fact that we must control all $N $ duplicates at once. 

In this paper we show that a fundamental protocol in quantum information, quantum teleportation, can be performed using the encoding (\ref{becqubit}). That is, an unknown state $ | \alpha, \beta \rangle \rangle $ at one location may be transferred onto another BEC qubit at a different location by the use of shared entanglement. The parameters $ \alpha, \beta $ correspond to spin coherent states that may be located at rather different regions of the Bloch sphere, such that the CV teleportation protocol cannot be applied.  Despite the large number of degrees of freedom in the Hilbert space of the spin coherent state, this is achieved in a fixed number of operations. We analyze the error of the protocol with the number of particles $ N $ in the spin coherent state under decoherence and find that it scales favorably with $ N $. We compare our protocol with an all-classical strategy and show that our protocol outperforms it, and achieves fidelities that are close to 100 \% for large $ N $.

\section{The teleportation protocol}
\label{sec:teleport}

We continue the tradition and call the heroes of our protocol as Alice (sender) and Bob (receiver). In order to make our scheme realistically implementable in the lab, we assume that only the following operations are available to manipulate the BEC qubits: (i) Coherent spin rotations corresponding to applications of the Hamiltonians
$S^x=a^\dagger b+b^\dagger a,$ $S^y=-ia^\dagger b+ib^\dagger a,$ and $S^z=a^\dagger a-b^\dagger b$; (ii) Projective measurements which collapse the coherent state
onto the number basis $|k \rangle=\frac{1}{\sqrt{k!(N-k)!}}(a^\dagger)^{k}(b^\dagger)^{N-k}|0 \rangle $; (iii) Two BEC 
qubit interactions corresponding to $ S^z_1 S^z_2 $. For details of the most likely experimental realization suitable to this protocol see works such as \cite{byrnes12,pyrkov13,abdelrahman14}.  Spin rotations corresponding to (i) have been performed in many contexts, and thus are possible using current 
experimental techniques \cite{riedel10, bohi09}. Projective measurements (ii) can be performed using either absorption or fluorescent imaging \cite{bucker09, andrews97b, depue00}. An alternative is to use dispersive measurements such as phase contrast imaging in the strong measurement regime to complete dephase the BEC in the number basis \cite{ilookeke14}. Although the two BEC qubit interaction has not been realized experimentally yet, 
this could either be done via cavity coupling methods \cite{byrnes12}, coupling by optical fiber \cite{pyrkov13}, geometric phase gates \cite{hussain14}, measurement based schemes in optical lattices \cite{cattani13}, or state dependent cold collisions \cite{treutlein06b}. 
The realization of strong coupling of a BEC to a cavity \cite{colombe07} and entanglement between a BEC and an atom \cite{lettner11} would suggest that such an interaction is within experimental reach.

Our quantum teleportation protocol for BEC qubit states is as follows. Alice is in possession of BEC qubits 1 and 2, and Bob has BEC qubit 3. BEC qubit 1 is in an unknown state (\ref{becqubit}) on the surface of the Bloch sphere which we parametrize $ \alpha = \cos \frac{\theta}{2} e^{-i \phi/2} , \beta = \sin \frac{\theta}{2} e^{i \phi/2} $, with $ \theta=[0,\pi], \phi = [-\pi,\pi] $. For simplicity we assume that $ \theta = \pi/2 $, such that the initial state is along the equator of the Bloch sphere. Teleporting a BEC qubit for arbitrary state on the Bloch sphere can be performed by adapting
the method for the two coordinates $ (\theta,\phi) $ \cite{pyrkov14}. The initial state of whole system is
\begin{equation}
\label{initial}
|\frac{e^{-i\phi/2}}{\sqrt2},\frac{e^{i\phi/2}}{\sqrt2}\rangle\rangle_1 
|\frac{1}{\sqrt2},\frac{1}{\sqrt2}\rangle\rangle_2
|\frac{1}{\sqrt2},\frac{1}{\sqrt2}\rangle\rangle_3 .
\end{equation}
The protocol then follows the following sequence: 1) Apply the entangling gate $ S^z_2 S^z_3 $ on BEC qubits 2 and 3
for a time $ T = 1/\sqrt{2N} $; 2) Apply the entangling gate $ -S^z_1 S^z_2 $ on BEC qubits 1 and 2 for a time $ \tau = 1/\sqrt{2N} $; 3) Apply a Hadamard gate on BEC qubit 1; 4) Measure BEC qubits 1 and 2 in the $ | k \rangle $ basis; 5) Classically transmit the binary result of whether the measurement outcome on BEC qubit 1 is $ k_1 < N/2 $ or otherwise to Bob; 6) Rotate BEC qubit 3 by an angle $ \pi $ if $ k_1 < N/2 $. At this point Bob now has possession of Alice's BEC qubit to a good approximation completing the teleportation.

We now explain how the protocol works. After the first entangling gate the state is
\begin{equation}
e^{-iS^z_2 S^z_3 T}|\frac{1}{\sqrt2},\frac{1}{\sqrt2}\rangle\rangle_2 |\frac{1}{\sqrt2},\frac{1}{\sqrt2}\rangle\rangle_3 
=\frac{1}{\sqrt{2^N}} \sum_{k_2=0}^N \sqrt{C_N^{k_2}} |k_2 \rangle | \frac{e^{-i(2 k_2-N)T}}{\sqrt{2}} , \frac{e^{i(2 k_2-N)T}}{\sqrt{2}} \rangle \rangle_3 , 
\label{entangling}
\end{equation}
where we have used the binomial representation the state of two BEC qubits. The state between BEC qubits 2 and 3 is discussed in Ref. \cite{byrnes12} in detail. This is analog of Bell state for coherent spin states in this representation. On application of the entangling gate for a time $ \tau $ between BEC qubits 1 and 2 we have
\begin{equation}
\frac{1}{\sqrt{2^N}} \sum_{k_2=0}^N \sqrt{C_N^{k_2}} | \frac{e^{-i(\phi/2-(2 k_2-N)\tau)}}{\sqrt{2}} , \frac{e^{i(\phi/2-(2 k_2-N)\tau)}}{\sqrt{2}} \rangle \rangle_1 |k_2 \rangle 
| \frac{e^{-i(2 k_2-N)T}}{\sqrt{2}} , \frac{e^{i(2 k_2-N)T}}{\sqrt{2}} \rangle \rangle_3 . 
\end{equation}
The Hadamard operation consists of making the replacement $ a \rightarrow (a+b)/\sqrt{2} , b \rightarrow (a-b)/\sqrt{2} $ on BEC qubit 
1.  This can be realized by a $ \pi/2 $ rotation around $ S^z $, a $ \pi/2 $ rotation around $ S^x $, followed by a $ \pi/2 $ rotation around $ S^z $ as realized using techniques described in Refs. \cite{abdelrahman14,bohi09}. This gives
\begin{equation}
\sum_{k_2=0}^N \sqrt{\frac{C_N^{k_2}}{2^N}} | \cos(\frac{\phi}{2}-(2 k_2-N)\tau) , -i\sin(\frac{\phi}{2}-(2 k_2-N)\tau) \rangle \rangle_1 
|k_2 \rangle | \frac{e^{-i(2 k_2-N)T}}{\sqrt{2}} , \frac{e^{i(2 k_2-N)T}}{\sqrt{2}} \rangle \rangle_3 , 
\end{equation}
Expanding BEC qubit 1 in the $ |k \rangle $ basis gives
\begin{multline}
\frac{1}{\sqrt{2^N}} \sum_{k_1,k_2=0}^N \sqrt{C_N^{k_1} C_N^{k_2}}\cos^{k_1}(\phi/2-(2 k_2-N)\tau) (-i)^{N-k_1} 
\sin^{N-k_1} (\phi/2-(2 k_2-N)\tau) 
\\
\times | k_1 k_2 \rangle | \frac{e^{-i(2 k_2-N)T}}{\sqrt{2}} , \frac{e^{i(2 k_2-N)T}}{\sqrt{2}} \rangle \rangle_3 .
\end{multline}
The resulting state of Bob's BEC qubit after the measurement on BEC qubits 1 and 2 is
\begin{align}
\label{bobsqubit}
| \frac{e^{-i(2 k_2-N)T}}{\sqrt{2}} , \frac{e^{i (2 k_2-N)T}}{\sqrt{2}} \rangle \rangle_3 .
\end{align}
Here, $ k_2 $ is the measurement outcome of measuring BEC qubit 2.

The probability distribution for the measurement in step 4 is
\begin{equation}
\label{probab}
p(k_1,k_2)=\frac{1}{2^N} C_N^{k_1} C_N^{k_2} \cos^{2k_1}(\phi/2-(2 k_2-N)\tau ) \sin^{2N-2k_1}(\phi/2-(2 k_2-N)\tau) .
\end{equation}
To obtain some insight into this expression let us use the approximation 
$ C_N^k \cos^{2k} x \sin^{2N-2k} x \approx A \exp[-2N (x \pm \frac{1}{2}\arccos (2k/N-1))^2 ] $ which is valid for $ N \gg 1 $ and $ x \in [-\pi/2,\pi/2] $, where the coefficient $ A \sim O(1) $. Using the binomial approximation $ \frac{C_N^k}{2^N} \approx \sqrt{\frac{2}{\pi N}}\exp [-\frac{N}{2} (2 k/N-1)^2 ] $, we can then write
\begin{multline}
\label{approx}
p(k_1,k_2) \propto 
\exp \left[- \frac{1}{2} \left( \frac{2 k_2 -N}{\sqrt{N}} \right)^2 \right] \\
\times \exp \left[ - \left( 2k_2-N - \sqrt{\frac{N}{2}}( \phi \pm \arccos(2k_1/N-1)) \right)^2 \right] .
\end{multline}
\begin{figure}
\begin{center}
\scalebox{0.35}{\includegraphics{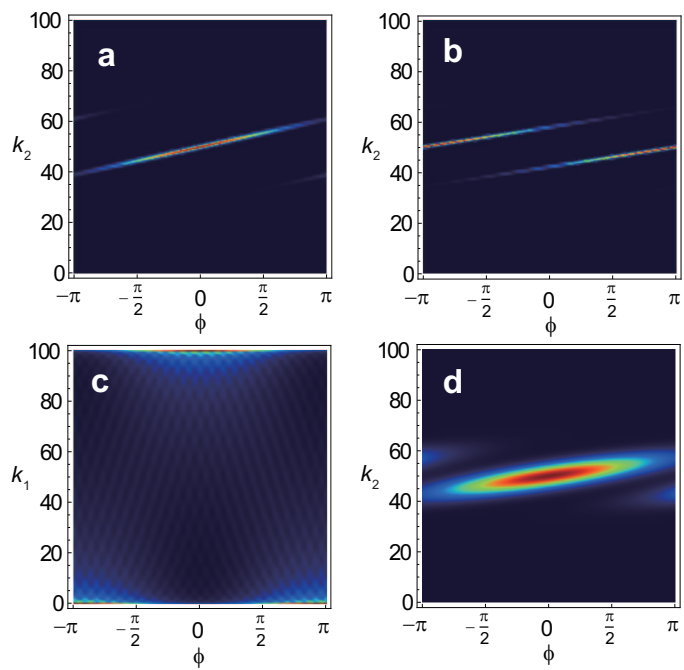}}
\caption{
Probability distributions of measurements on Alice's BEC qubit. Probability $ p(k_1,k_2) $ for (a) $ k_1 = N $, no dephasing ($ \gamma = 0 $); (b) $ k_1 = 0 $, no dephasing ($ \gamma = 0 $); (c) Total probability $ \sum_{k_2} p(k_1,k_2) $ with $ \gamma = 0 $; (d) $ k_1 = N $, with dephasing ($ \gamma = 2 $). All calculations use $ N = 100 $. 
\label{fig1} }
\end{center}
\end{figure}
The first factor in the probability distribution states that $ \langle k_2 \rangle = N/2 $ and
random variations to this occur with standard deviation $ \sqrt{N}/2 $. The second factor involves
a correlation between Alice's angle $ \phi $ to be teleported and $ k_2 $. This 
correlation gives rise to the ability to teleport $ \phi $, because after the measurement 
Bob holds the state (\ref{bobsqubit}), which must be correlated to the original state. 
The probability distribution $p(k_1,k_2)$ has a maximum when
\begin{equation}
\label{linearfunc}
2k_2-N= \sqrt{\frac{N}{2}} \phi \mp \sqrt{\frac{N}{2}}\arccos(2k_1/N-1) .
\end{equation} 
Distributions of $p(k_1,k_2)$ at fixed $k_1$ versus $\phi$ and $k_2$ are presented on Figure \ref{fig1}. We have sharp correlations between $\phi$ and $k_2$ which obey the linear function (\ref{linearfunc}). This arises due to the standard deviations of the second term in (\ref{approx})
being $ \sim O(1) $, thanks to the choice $ \tau = 1/\sqrt{2N} $.

In order to match the state of Bob's BEC qubit (\ref{bobsqubit}) to the initial state, by substituting (\ref{linearfunc}) into (\ref{bobsqubit}) we see that the entangling time $ T $ should be set to $T=1/\sqrt{2N}$, as given in our prescription. The state of 
Bob's BEC qubit at the end of step 4 is then with probability approaching unity for large $ N $
\begin{align}
| \frac{e^{-i[\phi/2 \pm \frac{1}{2}\arccos(2k_1/N-1) ]}}{\sqrt{2}}, \frac{e^{i[\phi/2 \pm \frac{1}{2}\arccos(2k_1/N-1) ]}}{\sqrt{2}} \rangle \rangle_3
\label{bobfin}
\end{align}
This is equal to the initial state, up to some known angle. 
The $ \pm \frac{1}{2}\arccos(2k_1/N-1)$ term is a random offset to $ \phi $ which must be classically corrected in order to match Bob's BEC qubit to Alice's. The probability distribution of $ k_1 $ is shown 
in Figure \ref{fig1}c. This shows a broad distribution which is mostly peaked at $ k_1 = 0 $ or $ N $ depending on which hemisphere $ \phi $ is in. Thus although the sign of the correction term $ \arccos(2k_1/N-1)$ is random, in these cases the correction to $ \phi $ is either $ \pm 0 $ or $ \pm \pi $ making it invariant of the sign. For other measurement outcomes the $ \pm $ makes the correction to the angle is imperfect, and a simple correction recipe is to take the midpoint and add an angle $ \pi $ to $ \phi $ if $ k_1 < N/2 $ and $ 0 $ otherwise. This means only a binary variable needs to be sent to Bob. To minimize the imperfect correction cases we may introduce a cutoff $ k_1^{\mbox{\tiny cut}} $ to the measurement results of BEC qubit 1 where the outcomes 
$ k_1^{\mbox{\tiny cut}} < k_1 < N - k_1^{\mbox{\tiny cut}} $ are discarded. The introduction of this parameter makes our protocol a conditional scheme, where 
in exchange of a reduced success probability, the fidelity of Bob's state with respect to Alice's state is improved. 
In terms of standard qubit teleportation, the $ k_1 $ measurement plays the role of the outcome of the Bell measurement,
which must be classically communicated to Bob in order to make the ``corrections'' to the final state \cite{bennett93}. Performing these corrections in Step 6 completes the teleportation protocol, and to a good approximation Bob is now in possession of Alice's initial state. 
\begin{figure}
\begin{center}
\scalebox{0.5}{\includegraphics{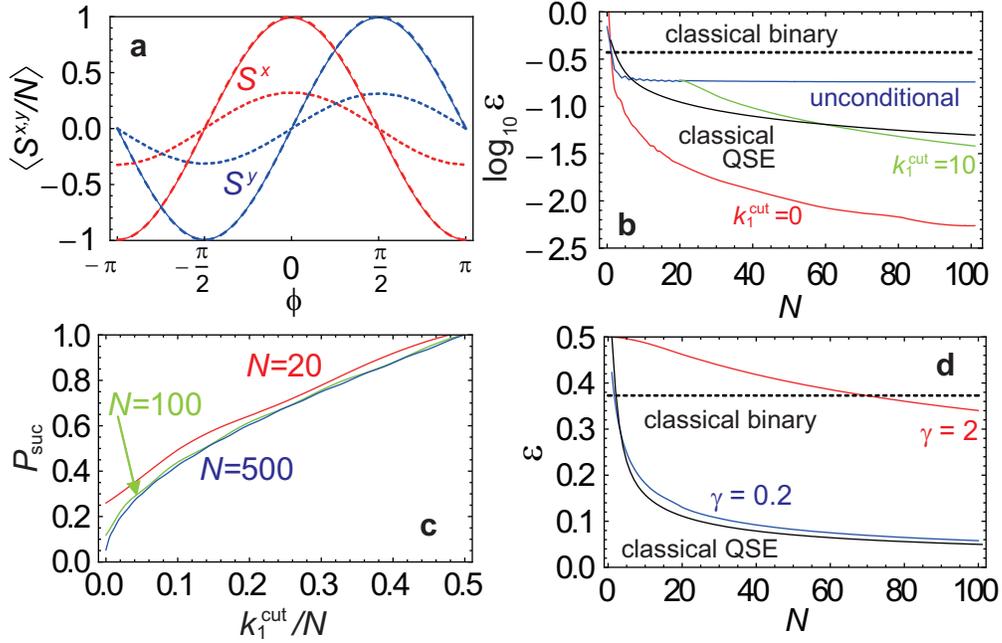}}
\caption{\label{fig2}
Performance of the proposed teleportation scheme. (a) Average spin of Bob's BEC qubit with (dotted lines) and without dephasing (solid lines) for $ N = 100 $ and the times of the entangling gate $ T =\tau = 1/\sqrt{2N} $. The ideal case of Alice's original state with $ \langle S^x \rangle/ N = \cos \phi $, $ \langle S^y \rangle/ N = \sin \phi $ is shown for comparison (dashed lines).  Parameters used are $ k_1^{\mbox{\tiny cut}} = 0$ and $ \gamma = 2 $ for all calculations including dephasing. (b) Error of teleported BEC qubit as measured by the trace distance in the Bloch sphere averaged for all $ \phi $ with boson number $ N $ for the conditional case of $ k_1^{\mbox{\tiny cut}} =0,10 $, and the unconditional case where all $ k_1 $ measurements are kept with no dephasing $ \gamma = 0 $. The classical bounds using quantum state estimation (QSE) in Ref. \cite{derka98} and Eq. (\ref{clasbin}), based on binary variable estimation, is shown for comparison.  (c) Success probability of the conditional protocol  $ P_{\mbox{\tiny suc}}  $ for $\phi = 0 $ and the boson numbers as shown.  (d) Error of teleported BEC qubit as measured by the trace distance including dephasing for the parameters shown and $ k_1^{\mbox{\tiny cut}} =0 $. The classical bounds of Ref. \cite{derka98} and Eq. (\ref{clasbin}) are shown for comparison. }
\end{center}
\end{figure}

\section{Comparison to an all-classical strategy and decoherence}

We now discuss the performance and quantumness of our teleportation scheme. Figure \ref{fig2}a shows the performance of the teleportation scheme. For perfect teleportation, we would expect that Alice's state is transferred to Bob with expectation values $ \langle S^x \rangle/ N = \cos \phi $, $ \langle S^y \rangle/ N = \sin \phi $.  Figure \ref{fig2}a shows that state of Bob's BEC after the teleportation protocol described in the previous section is virtually identical to the ideal case when no dephasing is present.   An important issue when discussing quantum teleportation is evaluating the ``quantumness''
of the protocol \cite{braunstein05,braunstein00,braunstein01}. That is, in comparison to an purely classical strategy of transmitting 
information, what improvement does the quantum protocol achieve in the fidelity of transfer of the original
state? Such bounds are useful in experimentally verifying that a quantum teleportation has indeed taken
place, and places a boundary between classical and quantum communication. The problem of estimating the quantum state of $ N $ copies of 
a qubit on the equator of the Bloch sphere has been analyzed in Ref. \cite{derka98}. Since our aim is to transfer a spin coherent state, which is nothing but $ N $ copies of a qubit, we may directly use this result. It is in our case more convenient to compare the trace distance 
\begin{align}
\varepsilon = \frac{1}{2} \sqrt{ (\langle S^x_{\mbox{\tiny Bob}} \rangle/N - \cos \phi)^2 + (\langle S^y_{\mbox{\tiny Bob}} \rangle/N - \sin \phi)^2 + (\langle S^z_{\mbox{\tiny Bob}} \rangle/N)^2 } 
\end{align}
rather than fidelities. 
In Figure \ref{fig2}b we compare the classical and quantum strategies. We see that for the zero dephasing case and $ k_1^{\mbox{\tiny cut}} =0 $, the protocol presented here outperforms the classical boundary, showing a quantum enhanced state transfer using entanglement. The superior performance of the quantum protocol can be understood qualitatively in the following way. In the classical strategy, the location of the initial state on the Bloch sphere can be estimated with standard deviation $ S^j/N \sim 1/\sqrt{N} $. Meanwhile, the sharp correlations in the second term of (\ref{approx}) give variations of order $ S^j/N \sim 1/N $, which allows for smaller fluctuations of the teleported state, giving rise to a lower error $ \varepsilon $. The $ k_1^{\mbox{\tiny cut}} =0$ is the optimal conditional case for our teleportation scheme and for the unconditional case (no $ k_1^{\mbox{\tiny cut}} $ parameter) the error of the teleportation scheme is due to imperfect classical correction of Bob's final state. 

The success probability of the conditional case may be quantified according to 
\begin{align}
P_{\mbox{\tiny suc}} = \sum_{k_2=0}^N  \sum_{
{k_1\le k_1^{\mbox{\tiny cut}} \atop k_1\ge N-k_1^{\mbox{\tiny cut}}}} p(k_1,k_2) 
\end{align}
and is plotted in Fig. \ref{fig2}c for $ \phi=0 $.  Other choices of $ \phi $ give little dependence to the curves, hence the values plotted here are representative of any state to be teleported. We see that for $ k_1^{\mbox{\tiny cut}} $ normalized to $ N $ the curves quickly converge to a universal curve as $N $ is increased.  Thus for a $ k_1^{\mbox{\tiny cut}}/N $ we may estimate the success probability using Fig. \ref{fig2}c.  However, the $ k_1^{\mbox{\tiny cut}} $ required to beat the classical bound depends on $ N $ as can be seen in Fig. \ref{fig2}b, which in turn affects the success probability.  For example, for $ N =100 $ a threshold of $ k_1^{\mbox{\tiny cut}} \approx 10 $ is required to beat the classical bound, which has $ P_{\mbox{\tiny suc}} \approx 0.4 $.  

While beating classical bounds based on quantum state estimation is important to show under certain circumstances, we point out that even the unconditional case performs a non-trivial task. One remarkable aspect of the protocol is that only a binary variable is classically sent between Alice and Bob, yet a spin coherent state involving a macroscopic number of spins is transferred between the two parties. Hence a continuous parameter $ \phi $ is transferred (within a finite fidelity) using only a binary variable.  For a binary variable that contains information of which hemisphere $ \phi $ is located in, the average expected error may be calculated as 
\begin{align}
\epsilon_{\mbox{\tiny bin}} & = \frac{1}{\pi} \int_{-\pi/2}^{\pi/2} d \phi \frac{1}{2} \sqrt{ (1-\cos \phi)^2 + \sin^2 \phi } \nonumber \\
 & = \frac{4-2\sqrt{2}}{\pi} \approx 0.37 .
\label{clasbin}
\end{align}
This average error is plotted on Figs. \ref{fig2}b and \ref{fig2}d.  In terms of the information exchanged between Alice and Bob, the unconditional protocol may be seen to outperform the average expected error, which can only be possible by the use of entanglement as a shared resource. 

In any experimental situation decoherence is inevitably present, and it is an important question how well the teleportation can be carried out under such conditions. Typically, 
quantum effects such as entanglement, which teleportation is based on, disappear at a timescale extremely quickly for macroscopic objects. Remarkably for the entangled states that we consider in this protocol, this is not the case. We consider decoherence in the form of a generic dephasing process, which can be described by the master
equation
\begin{align}
\frac{d \rho(t)}{dt} = -i [H(t), \rho] - \frac{\gamma}{2} \sum_{i=1}^3 \left[ (S^z_n)^2 \rho 
-2 S^z_n \rho S^z_n + \rho (S^z_n)^2 \right] ,
\label{dephasingmaster}
\end{align}
where $ H(t) $ is the Hamiltonian sequence described above. For concreteness we assume that the system starts in the initial state (\ref{initial}), and the dephasing 
takes place for the duration of the entangling gates (Steps 1 and 2). We do not consider dephasing to 
take place during the Hadamard step since this can be combined with the measurement step as a projection on the $ x $ basis. Figure \ref{fig1}d shows the effect of the dephasing term on the probability distribution $ p(k_1,k_2) $. We see that the sharp correlation between $ \phi $ and $ k_2 $ is reduced and the distribution is broadened due to the dephasing, due to random variations in the correlations between BEC qubit 1 and 2. Figure \ref{fig2}a shows the output state at Bob's BEC qubit which shows a degradation of the fidelity from the ideal zero decoherence case. The effect of the decoherence can be seen to diminish the amplitude of the output spins due to a combination of degradation of the correlations between Alice and Bob's states, as well as a direct decoherence on Bob's qubit.

Figure \ref{fig2}d shows the scaling of the error under decoherence with $ N $. Interestingly, we see that the case including dephasing tends to improve with $N $. 
There are several contributing factors to this. The first reason is that the fidelity of the protocol itself improves with $ N $, as can be observed from the ideal $ \gamma = 0 $ case. A second reason is due to the fact that as $ N $ grows, the gate times that are necessary actually decreases as $ \tau,T = 1/\sqrt{2N} $. This means that there is less time for the dephasing to destroy the states for large $ N $, due to the shorter times that the decoherence acts. A third reason is that in our protocol we never use states that are extremely sensitive to decoherence. Decoherence is intrinsically a state-dependent process, and the time that quantum coherence can survive depends upon the encoding. For example, a qubit that is encoded on a Schrodinger cat state $ \alpha | 1,0 \rangle \rangle + \beta | 0,1 \rangle \rangle$ dephases to a mixed state under (\ref{dephasingmaster}) in time $ \sim 1/(N^2 \gamma) $, whereas the spin coherent state dephases in time $ \sim 1/\gamma $ \cite{hecht,byrnes13}. Thus although such 
cat states could in principle be used to encode quantum information for teleportation, from a decoherence point of view this is generally not a wise choice. Thus the use of the entangled states corresponding to times $ T = \tau = 1/\sqrt{2N} $ can be seen as the key to allowing teleportation to be performed without the very fast degradation of coherence that cat states would suffer. We estimate the a dephasing rate of $ \gamma<0.2 $ is sufficient to beat the classical bound, a value which should be possible according to the estimates in Ref. \cite{pyrkov13}.  

We note that finite measurement resolution can also affect the fidelity of the protocol.  The unconditional scheme is rather insensitive to measurement errors as the only readout that is required is to determine whether $ k_1 < N/2 $ or the reverse. This follows from the fact that most of the distribution is peaked around $ k_1 = 0,N $ (see Fig. \ref{fig1}c) discriminating between these should be performed accurately.  The conditional scheme with the $ k_1^{\mbox{\tiny cut}} $ parameter is more sensitive to measurement errors as it requires discarding unfavorable $ k_1 $ outcomes. However, techniques to improve the measurement resolution have been developed to nearly the single atom level \cite{bucker09} which would benefit the current proposal.

\section{Conclusions}
\label{sec:conc}
In summary, we have introduced a protocol that can teleport a spin coherent state lying on an arbitrary position on the Bloch sphere 
between two parties. A classical binary random variable needs to be sent in order to ``correct'' the transmitted state. 
Although we have used the language of BEC qubits, the algebra used is that of spin coherent states, so in principle should apply to 
other systems, such as atomic ensembles as the only operations that are used are the total spin operators and projective measurements. 
The primary novel aspect of the protocol is that spin coherent states on a larger region of the Bloch sphere can be teleported, beyond the standard continuous variables approximation based on canonical position and momentum variables. This can be achieved with asymptotically unit fidelity particularly for large boson number systems. Furthermore, decoherence scales favorably with $ N $ due to the use of entangled states with interaction times $ \tau, T \sim 1/\sqrt{N} $, such that it should be applicable to realistically sized systems. Beating the classical bound becomes harder for large $ N $ as it is possible with increasing accuracy to estimate Alice's state and classically communicate this to Bob.  However, even if classical bounds are not exceeded, a non-trivial task is being performed by the protocol as the classical binary variable exchanged between Alice and Bob which clearly cannot contain complete information of $ \phi $ itself. In the classical protocol, an estimate of $ \phi $ is directly transferred, and has a fundamentally different character to the proposed scheme. 

In standard continuous variables teleportation, the fidelity reaches unity for an infinitely squeezed state.  An infinitely squeezed state is unphysical in the sense that it has an infinite number of photons \cite{braunstein05}. Therefore in practice there are physical limits to the fidelity that can be achieved by such an approach, with the best experimental value to date being 83\% \cite{yukawa08,andersen13}.  In our protocol the entanglement resource is a $ S^z S^z $ interaction, which in terms of energy does not produce any excitations outside of the ground states of the atoms in the BEC.  While various decoherence channels exist for generating this type of entanglement \cite{pyrkov13}, we do not expect that there is a limitation in terms of unphysical states in the same way as the continuous variable teleportation case.  In this way our approach may be a complementary approach to standard continuous variable techniques that has the possibility of reaching higher fidelities .

\ack
T. B. thanks Eugene Polzik, Mikhail Lukin, and Ujjwal Sen for discussions. This work is supported by the Transdisciplinary Research Integration Center, Japan Russia Youth Exchange Center, RFBR grant no. 14-07-31305, the Okawa Foundation, the Inamori Foundation, NTT Basic Laboratories, and JSPS KAKENHI Grant Number 26790061.

\end{document}